\documentclass[preprint,longbibliography]{revtex4-1}
\usepackage{graphicx}  
\usepackage{bm}     
\usepackage{amssymb}   
\usepackage{amsmath}
\usepackage{upgreek}

\begin{document}

\author{Jon Alm Eriksen} \email{jonaerik@fys.uio.no}
\affiliation{Department of Physics, University of
  Oslo, P. O. Box 1048 Blindern, N-0316 Oslo, Norway}
\affiliation{Institut de
  Physique du Globe de Strasbourg, University of Strasbourg/EOST, CNRS, 5 rue
  Descartes, F-67084 Strasbourg Cedex, France} 
\author{Renaud Toussaint}
\affiliation{Institut de Physique du Globe de Strasbourg, University of
  Strasbourg/EOST, CNRS, 5 rue Descartes, F-67084 Strasbourg Cedex, France}
\author{Knut J\o rgen M\aa l\o y} \affiliation{Department of Physics,
  University of Oslo, P. O. Box 1048 Blindern, N-0316 Oslo, Norway}
\author{Eirik Flekk\o y} \affiliation{Department of Physics, University of
  Oslo, P. O. Box 1048 Blindern, N-0316 Oslo, Norway} \author{Bj\o rnar
  Sandnes} \affiliation{College of Engineering, Swansea University, Bay Campus,
  Fabian Way, SA1 8EN Swansea, UK}

\title{Pattern formation of frictional fingers in a gravitational potential}
\date{\today}

\begin{abstract}
  Aligned finger structures, with a characteristic width, emerge during the
  slow drainage of a liquid/granular mixture in a tilted Hele-Shaw cell. A
  transition from vertical to horizontal alignment of the finger structures is
  observed as the tilting angle and the granular density are varied. An
  analytical model is presented, demonstrating that the alignment properties is
  the result of the competition between fluctuating granular stresses and the
  hydrostatic pressure. The dynamics is reproduced in simulations. We also show
  how the system explains patterns observed in nature, created during the early
  stages of a dyke formation.
\end{abstract}
\maketitle

\section{Introduction}
Subsurface flows tend to converge on high-conductivity pathways such as rock
fractures, joints and faults. Flow of oil and gas in fractured reservoirs,
groundwater transport, magma flow and pollutant transport in fractured porous
media are therefore often dominated by the interactions between the flowing
fluids, the confining geometries, and granular rock fragments residing in the
cracks or faults.

A range of flow patterns can emerge when one fluid displaces another fluid in
such confined spaces~\cite{sahimi2011flow}. These flow patterns are caused by
the interplay between different stabilizing and destabilizing effects, like
surface tension, gravity, pore size fluctuations, wettability properties and
granular effects. Viscous fingering is a well-known example of a fluid flow
instability. An initially straight interface between two immiscible fluids of
different viscosities develops undulations that grow to form fingers when the
less viscous fluid invades the more viscous host fluid
\cite{saffman_taylor58,bensimon_rmp1986}. In rough fractures or a porous
medium, disorder in the form of variations in pore sizes perturbs the invading
interface, generating fractal two-phase flow structures with no intrinsic
length scale~\cite{lenormand1985,maaloy1985prl,
  chen1985,toussait2005influence}.

Gravity has a profound effect on the flow patterning in situations where a
density difference between the fluids exists, and where the flow geometry is
not strictly horizontal. For example, in density driven convection, the
interface between a dense fluid overlying a less dense fluid becomes unstable,
with dense fluid fingers sinking and low density fingers rising (the
Rayleigh-Taylor instability \cite{taylor1950}). With the less dense fluid on
top on the other hand, the hydrostatic pressure stabilizes the interface at a
given height. During slow drainage of a porous medium, a competition exists
between the stabilizing effect of gravity, and the pore scale disorder that
increases the roughness of the invasion front
\cite{birovljev1991,meheust2002interface}.

Rock fractures and other high permeability flow paths can be filled with
granular debris and fault gouge from cataclastic processes and
erosion~\cite{amitrano2002fracture,nguyen2009energetics}, or materials carried
by fluid flow. Multiphase flows involving both a combination of different
fluids and a loose packing of granular materials have proved a particularly
rich vein of pattern formation as frictional fluid dynamics is added to the
well-known two-phase flow mechanisms~\cite{sandnes2011natcomm}. Recently
observed flow patterning processes include multiphase fracturing of deformable
granular packings~\cite{chevalier2009morphodynamics, shin2010fluid,
  varas2011venting,holtzman2012capillary,sandnes2011natcomm,eriksen2015invasion,
  sandnes2015gasmigration,niebling2012dynamic}, decompaction
fingers~\cite{johnsen2006pre, johnsen2008decompaction}, frictional
fingers~\cite{sandnes2007prl,knudsen2008pre} and bubble formation
\cite{sandnes2011natcomm,eriksen2015bubbles}.

Here we study the stabilizing effect of gravity on a patterning flow, as air
displaces a liquid/granular mixture during drainage of a Hele-Shaw cell at
shallow tilt angles. The receding interface accumulates a front of granular
material, and an instability caused by a competition between capillary and
frictional forces results in an emerging pattern of frictional fingers --
canals of air separated by branches of compacted grains
\cite{sandnes2007prl,knudsen2008pre,sandnes2011natcomm}. The symmetry breaking
by gravity on the tilted system causes a stabilization of the drainage front
and a resulting directionality and alignment of the finger structures.

We find that the key to the finger alignment direction is a competition between
gravity and fluctuations of the inter-granular stresses. Analogous to drainage
in porous media \cite{meheust2002interface}, random fluctuations in threshold
pressures cause a disruption of the stabilizing effect of gravity. However,
unlike porous media, there is in our system a spontaneous emergence of a {\em
  characteristic length}, the finger width, $2\Lambda$ ($\Lambda$ denotes half
the finger width). The magnitude of the disruption of the invasion front
becomes a relative quantity with respect to this length scale. We show that the
basic assumption that the effective granular friction stresses at the interface
arises as a sum of a set of uncorrelated random contributions, is sufficient to
give a theoretical prediction of the transition between the different pattern
morphologies.

We also show how the pattern forming mechanism provides a new understanding of
the small-scale flow properties during magmatic dyke formations, i.e.~the
penetration of a sheet of magma into a fracture of a pre-existing rock body.
The small-scale flow properties during this formation, when magma interacts
with the host rock, is largely unknown~\cite{rivalta2015review}, as the
formation occurs deep beneath the Earth's crust. Rock faces in the Israeli
desert \cite{baer1987, baer1991} display aligned finger structures which were
formed during a dyke formation. The structures have previously been attributed
to viscous fingers, due to the Saffman-Taylor
instability~\cite{saffman_taylor58}, between the fluidized host rock and a less
viscous dyke-related fluid in front of the invading magma~\cite{baer1991}. We
hypothesize here, that intergranular frictional forces between quartz grains in
the fluidized host rock, and not viscous forces of the fluids, govern the
formation of the pattern.

\section{The Experiment}

\begin{figure}
\includegraphics[width=8.6cm]{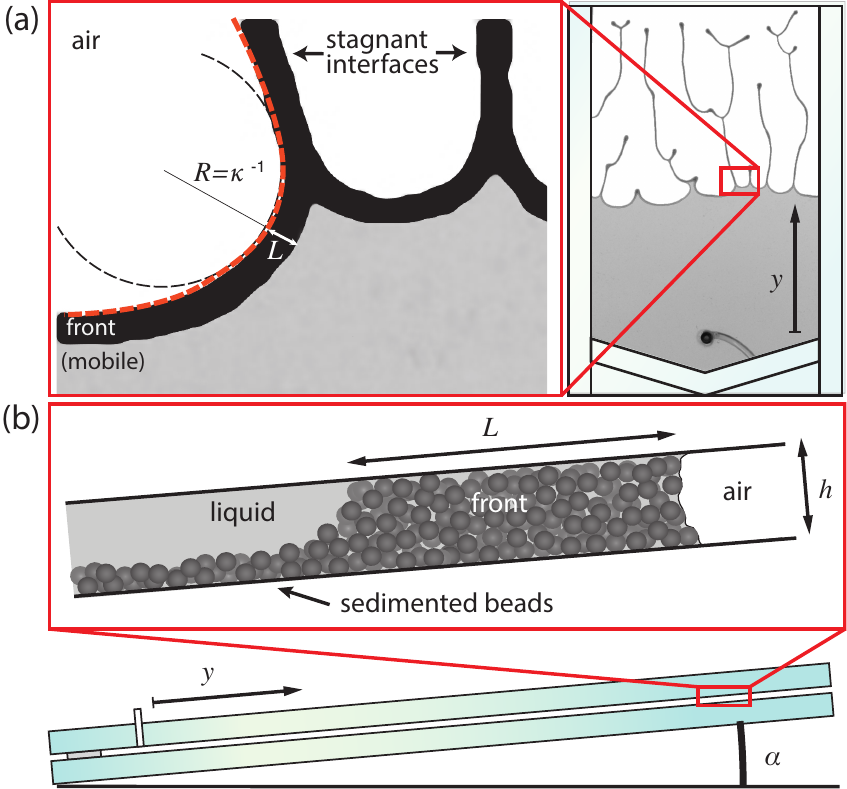}
\caption{(Color online) (a) Top view of the Hele-Shaw cell. The coordinate $y$
  is running from the outlet towards the upper edge of the cell, $\kappa$ is
  the curvature (inverse of the in-plane radius of curvature $R$) along the
  interface (orange dashed line). The front is a region of accumulated grains
  along the air-liquid interface; $L$ is the thickness of this front. The cell
  is 20$\times$30~cm$^2$. (b) Side view. The cell is tilted by an angle
  $\alpha$. The cell gap is $h=0.5$~mm. The filling fraction $\phi$ is the
  height of the initial sedimented granular layer relative to $h$.}
\label{fig:setup}
\end{figure}

Consider a rectangular 200 $\times$ 300~mm$^2$ Hele-Shaw cell with a gap
spacing $h=0.5$~mm [Fig.~\ref{fig:setup} (a)]. The cell is sealed along the
sides and base; the upper end is open to the ambient air. In preparation for
the experiment a granular material suspended in a 50\% (by volume) water-glycerol
mixture is injected into the horizontal cell through an inlet/outlet hole close
to the base of the cell. Excess mixture spills through the open edge such that
the granular suspension fills the entire cell. The granular
material---spherical glass beads with mean diameter
80$\pm10$~$\upmu$m---settles out of suspension, forming a layer of grains 
resting on the lower glass plate of the cell. The height of this layer,
relative to the cell gap, is denoted $\phi$, and quantifies the initial filling
fraction of the injected granular mixture relative to the random loose packing
fraction of the grains. The glass beads are polydisperse, and the variation in
size prevents crystallization of the sedimented bead packing. The density of
the glass beads and the water-glycerol mixture is $\rho_g=2.4$~g/cm$^3$ and
$\rho=1.13$~g/cm$^3$ respectively. The bead-fluid density contrast makes the
beads sediment on the bottom plate.

The long side of the cell is tilted by an angle $\alpha$ relative to the
horizontal plane [Fig.~\ref{fig:setup} (b)]. Here we report only results for
shallow tilt angles ($0^{\circ} \leq \alpha \leq 5^{\circ}$) where no sliding
of the granular layer takes place. The experiment commences by slowly draining
fluid from the outlet at the base at constant withdrawal rate
$q = 0.07$~ml/min, using a syringe pump (WPI, Aladdin 1000). The withdrawal
rate is slow enough to leave the layer of grains along the air/fluid interface undisturbed by the fluid
flow. As fluid is slowly drained, air starts to invade, and the meniscus along
the elevated open edge of the cell gradually recedes.  The system is imaged
from underneath using a PL-B742U Pixelink camera, and illuminated by a white
screen placed above. Compacted granular material appears dark in the images,
and empty regions of the cell appear white.

As the air displaces the liquid, grains accumulate along the air-liquid
interface and fill the cell gap, forming a dense pack in a region adjacent to the interface which we
refer to as the \emph{front} (see Fig.~\ref{fig:setup}). Only a small section
of the interface moves at any given time, and the motion consists of
incremental displacements, as the air fills an ever-increasing volume.  A
moving section tends to continue its motion over many consecutive increments
before it stops and the motion continues at another section. The interface
develops frictional fingers of air surrounded by a front~\cite{knudsen2008pre,
  sandnes2007prl, sandnes2011natcomm}, with a characteristic finger width. When
different fingers move towards each other, their fronts combine, and their
interfaces stagnate. The evolution continues until either the whole cell is
filled with air and stagnant fronts, or the air reaches the outlet. 

\begin{figure}
\includegraphics[width=8.6cm]{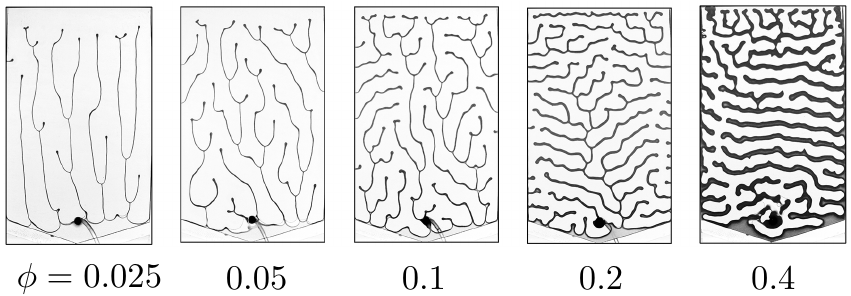}
\caption{Final configuration of the experimentally observed pattern at constant tilt angle,
  $\alpha=4^\circ$, with varying filling fraction $\phi$.  The finger alignment
  changes direction from vertical to horizontal as $\phi$ increases. Each image
  frame is 200~mm wide.}
\label{fig:tiltseries}
\end{figure}

When the cell is fixed horizontally ($\alpha=0^{\circ}$), the finger directions
are disordered and isotropic, and the resulting patterns are labyrinth
structures of stagnant fronts~\cite{knudsen2008pre, sandnes2007prl}.  When the
cell is tilted, the frictional fingers tend to
align~\cite{knudsen2009prelim_gravity}. The direction of alignment changes as
we vary $\alpha$ or $\phi$. Fig.~\ref{fig:tiltseries} shows the residual
patterns of granular material in the shape of narrow branches after 
all the grains have been packed at the end of each experiment. The figure displays results from a
series of experiments with increasing filling fraction, with the tilt angle
kept constant at $\alpha = 4^{\circ}$. The pattern of residual granular
material bears witness to the dynamics of the invasion process. At low $\phi$,
the air fingers march downwards, from top to bottom, leaving granular branches
aligned with the direction of gravity (vertical in the images). At high $\phi$,
the system makes a transition to sideways growing air fingers, leaving a trail
of horizontally aligned granular branches. Alternatively, by keeping $\phi$
constant, and increasing the tilt angle from $0^{\circ}$ to $5^{\circ}$, it is
possible to go from random labyrinthine pattern to horizontal alignment and
then to vertical alignment at high $\alpha$.

\begin{figure}
\includegraphics[width=8.6cm]{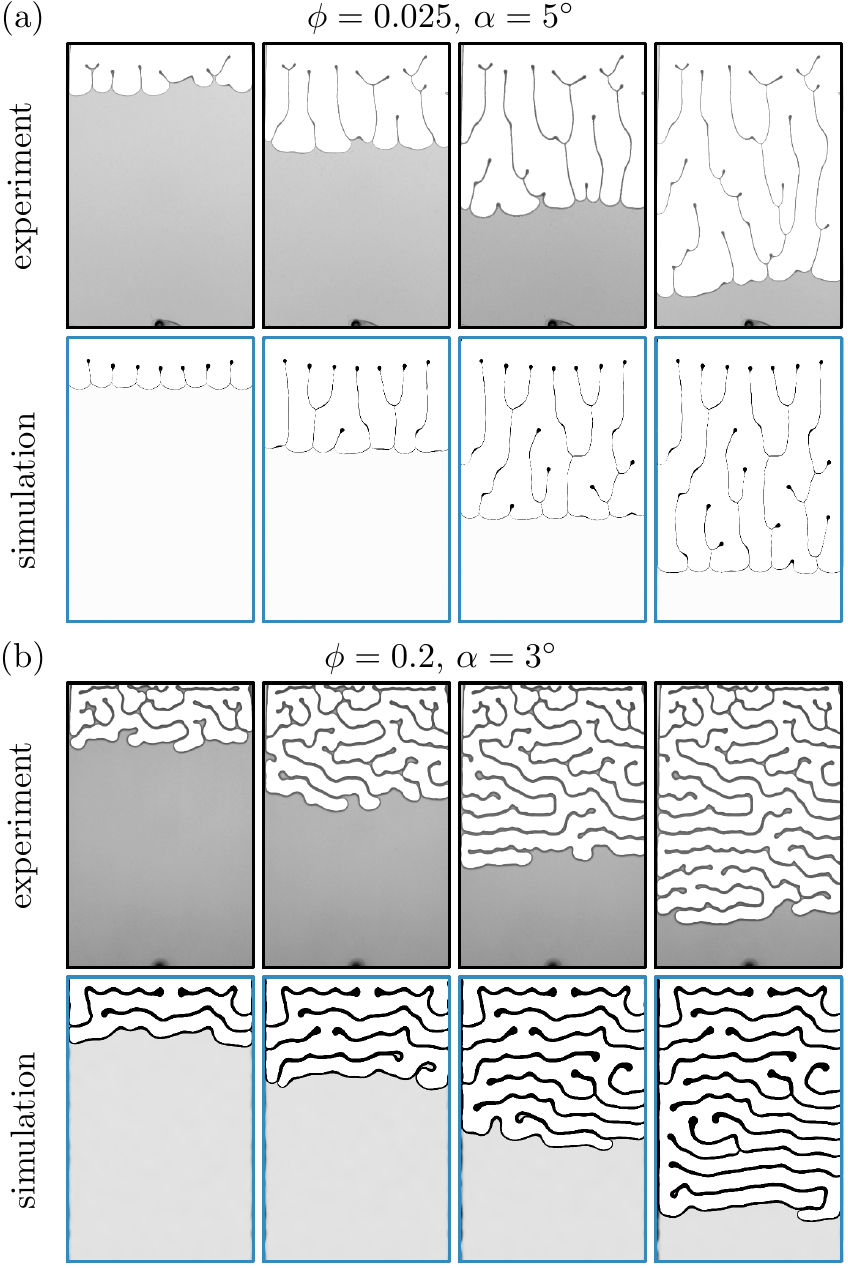}
\caption{(Color online) Snapshots of the dynamics, experiments versus
  simulations. (a) The pattern is dominated by vertically aligned fingers at
  $\phi=0.025$ and $\alpha=5^\circ$. (b) The pattern is dominated by
  horizontally aligned fingers at $\phi=0.2$ and $\alpha=3^\circ$. See (a) SM Video 1 and (b) SM Video 2~\cite{Eriksen_Supp}.}
\label{fig:dynamics}
\end{figure}

In the low $\phi$/high $\alpha$ range, hydrostatic height stabilization of the
receding interface dominates the dynamics, the fingers advance side-by-side
downwards, parallel to the gravitational field along the cell
[Fig.~\ref{fig:dynamics} (a), SM Video 1~\cite{Eriksen_Supp}]. Lateral growth
is inhibited by the presence of neighboring fingers on both sides; each finger
is confined to downwards growth. A finger will terminate its movement if it is bypassed and sealed off by its neighboring fingers. A finger can also split in two if a small region along the finger tip gets stuck, and each side of this region evolves to separate fingers. This typically happens when a finger tip widens, which seems to happen in conjunction with the termination of a neighboring finger. 
Finger termination and tip-splitting occur at
approximately equal frequencies [see Fig.~\ref{fig:dynamics} (a)]. We note that these patterns looks remarkably similar to
patterns generated when simulating retraction of a dewetting suspensions
\cite{thiele2009modelling}, although the setup is completely different.

\begin{figure*}
\includegraphics[width=\textwidth]{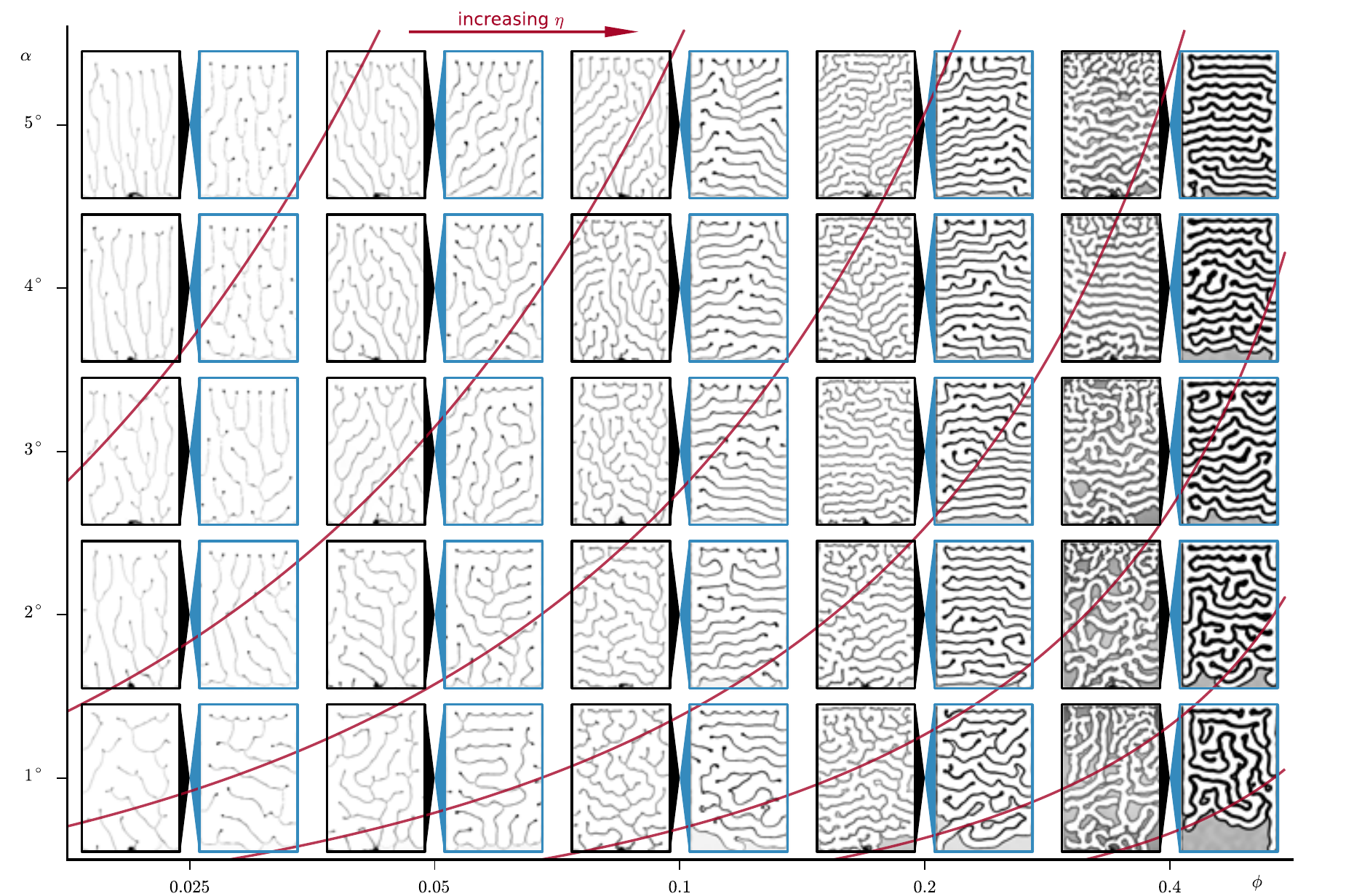}
\caption{(Color online) Pairwise comparison of the final configuration of
  experiments (black/left frames) to simulations (blue/right frames), for
  different values of the filling fraction ($\phi$), and the tilting angle
  ($\alpha$). The red lines indicate contours of constant $\eta$
   which are estimated up to a constant factor in Eq.~(\ref{eq:eta}). As $\eta$ increases, the vertical alignment turns into
  horizontal alignment, and then into no alignment. The value of $\eta$ doubles
  for every contour. The gravitational pull is pointing downwards in every
  frame.}
\label{fig:phasediagram}
\end{figure*}

As we increase $\phi$ and reduce $\alpha$, we observe a gradual transition in
the alignment; the fingers tend to grow with a directional component transverse
to the hydrostatic pressure gradient. In the intermediate range of $\phi$ and
$\alpha$, hydrostatic stabilization of the front occurs, but local pressure
fluctuations enables some fingers to get ahead. Sideways growth is preferred
for a finger that extends beyond its neighbors due to the hydrostatic pressure
gradient. The finger which manages to get ahead fills a larger fraction of the
horizontal direction, and advances layer by layer, creating a pattern of
horizontal lines (Fig.~\ref{fig:dynamics} (b), SM Video
2~\cite{Eriksen_Supp}). In the high $\phi$/low $\alpha$ range the local
pressure fluctuations dominate over the stabilizing effects, and alignment is
lost. A phase diagram of the alignment behavior of the end configurations is
shown in Fig.~\ref{fig:phasediagram}.

\section{Model}

As the dynamics are manifested by incremental movements of confined regions of
the interface, it is reasonable to assign a yield pressure threshold to every
point along the interface. When the pressure difference at the interface
exceeds the threshold at the weakest point along the interface, the
interface locally to that point deforms and moves a small step towards the
liquid phase. This approach has successfully modeled labyrinth patterns in a
similar setup \cite{sandnes2007prl,knudsen2008pre}, but without considering the
hydrostatic pressure differences induced by the tilting of the cell. In order
of quantify the yield pressure threshold, we will assign two local parameters
to the interface: the front thickness $L$ and the apparent in-plane curvature
$\kappa$.

The front thickness, $L$, is the distance from the air-liquid interface, in the
perpendicular direction, to the region of the liquid mixture where the beads no
longer fills the whole cell gap (see Fig.~\ref{fig:setup}). Note that the
packing of beads in the front remains in a static configuration before a
potential movement. We assign a yield stress $\sigma_Y(L)$ to every point
along the interface, which captures the static frictional properties of the
front. To be precise, $\sigma_Y$ is the yield stress acting normal to the
plane which approximate the air/liquid interface. This yield stress has
previously, in the context of labyrinth patterns \cite{knudsen2008pre} and of
plug formations in narrow tubes \cite{PhysRevLett.117.028002}, been assumed to
be exponentially increasing in front thickness $L$. The exponential behavior
can be justified by considering Janssen's model for stresses in packings of
grains, which assumes a linear relationship between the principal stresses in
the packing, in conjunction with the static Coloumb frictional stresses at the
plate boundaries of the cell. The yield stress may also have a curvature
dependence, as described in Ref.~\cite{eriksen2015bubbles}.  In the following,
we will, however, describe the yield stress as a linear function in $L$,

\begin{equation}
  \sigma_Y(L) = \frac{\sigma_\xi}{\xi} L,
\label{eq:sigmay}
\end{equation}

\noindent
for simplicity. The numerical comparison to the experimental behavior in the
subsequent section, will validate this approximation as sufficient for the
range of parameters that we consider here. The expression in the equation above
has two interpretations. We can interpret it as a linearization of a more
complicated function of $L$, e.g. the exponential behavior assumed in
\cite{knudsen2008pre, PhysRevLett.117.028002}. Alternatively, we can interpret
the yield stress as a sum of consecutive force bearing arc chains
\cite{jon2015numerical} which transmit frictional stresses, $\sigma_\xi$, from
the cell plates to the beads at the air-liquid interface. The characteristic
length of these chains is $\xi$, and the total number of chains scales with the
size of the front and therefore linearly in $L$.

The air-liquid surface tension at the interface acts at two different scales.
At the small scale, the interface makes bridges between wetting beads.  Each
point on a meniscus can be characterized by two principal radii of curvature.
By the Young--Laplace equation, the pressure drop over a meniscus is
proportional to the mean of the principal curvatures.  This means that in a
static configuration, each meniscus has the exact same mean curvature, up to
differences in the hydrostatic liquid potential, which we can ignore in a
horizontally oriented cell.

At a larger scale, we can identify a curvature which is averaged over several
neighboring beads. For our Hele-Shaw setup, the principle directions of the
average curvature are the in-plane and the out-of-plane directions with respect
to the cell plane.  We will disregard the curvature component in the
out-of-plane direction of the cell, i.e.~the curvature of the interface as it
is illustrated in the cross section in Fig.~\ref{fig:setup} b. The out of plane curvature is supposed roughly constant, i.e.~the surface
stress related to this component is constant along the in-plane direction of
the interface, and does, at our level of description, only contribute to a
constant global pressure drop. It plays no role when we later need to determine
the minimal yield stress.

The large scale surface behavior, i.e.~the surface behavior averaged over many
neighboring inter bead menisci, can be characterized by an effective surface
tension $\gamma$ \cite{knudsen2008pre}. The effective tension acts against the
increase of the apparent interface area during the displacement process, and
the associated pressure difference is simply $\gamma\kappa$.

We can now quantify the local yield pressure threshold. Let $\Delta p$ be the
difference between the air pressure, $p_\text{air}$, which is considered constant, and the
liquid pressure at the outlet of the cell, $p_\text{outlet}$. We assume that a section of the
interface is mobilized if

\begin{equation}
  \Delta p \geq \gamma\kappa + \frac{\sigma_\xi}{\xi} L - y \rho g \sin \alpha.
  \label{eq:threshold}
\end{equation}

\noindent
The first and second terms on the right hand side is the effective surface
stress and the yield stress [Eq.~(\ref{eq:sigmay})] described above. The
last remaining term is the hydrostatic pressure relative to the base of the
cell, $y$ is a coordinate running along the cell from the outlet, $g$ is the
gravitational acceleration and $\rho$ is the liquid density. 
This amounts to say that the local pressure in the fluid behind the meniscus, $p_\text{air} - \gamma \kappa$, is equal to the sum of the solid and the fluid stress there.
The fluid stress there is $p_\text{outlet} - y \rho g \sin \alpha$,
Hence, the solid stress there is 
$\sigma_{\perp\text{solid}} = p_\text{air}- (p_\text{outlet} - y \rho g \sin \alpha) - \gamma \kappa$
If the solid stress is equal or larger than $\sigma_\xi/\xi L$, the grain pack slides locally.
The pressure difference, $\Delta p$, will increase when the whole interface remains static
and liquid is drained from the system. The next moving section, at any given
time, is identified by local parameters $\kappa$, $L$ and $y$, which minimizes
the right hand side of Eq.~(\ref{eq:threshold}). As the section yields and
moves a small step towards the liquid, the local parameters are changed due to
the deformation and the accumulation of new beads onto the front.

\subsection{Numerical Validation}

We can reproduce the experimental behavior in a numerical simulation. The
numerical scheme has previously been used to simulate finger behavior in a flat
cell \cite{jon2015numerical}. We present here a summary of the numerical
strategy, and the modifications which are needed for the tilting of the
cell. Further details of the numerical scheme are described in
Ref.~\cite{jon2015numerical}.

The fluid interface (i.e. the boundary of the gas phase), can be represented as
a chain of nodes, labeled by an index $i$, where each node carries information
of the spatial coordinates $(x_i,y_i)$, and its nearest neighbors,
$i\pm1$. Such a chain can conveniently be implemented like a doubly linked
list. We couple this chain of nodes to a two dimensional mass field,
representing the grains. The complete filling of the cell gap, i.e.~the region
which constitutes the front, is indicated by the region of the mass field which
exceeds a threshold value. We make sure that the region of the mass field
adjacent to the chain, i.e.~the region of the front, exceeds this threshold in
the initial configuration of the system. The imposed dynamics described below
will maintain this state.

For each node we can identify the two local properties. First, the local front
length $L_i$ is represented as the shortest distance from any given node, to a
cell in the mass field which take a value below the threshold. This cell will
be referred to as the \emph{link} cell associated to the node. Second, we can
approximate the local curvature, $\kappa_i$, at node $i$, by numerical
differentiation of a spline approximation of the nearest and next nearest
neighbors $\{i,i\pm1, i\pm2\}$. By discretizing the right hand side of
Eq.~(\ref{eq:threshold}), we can now identify a pressure threshold $T_i$ for
each node,

\begin{equation}
T_i = \gamma\kappa_i + \frac{\sigma_\xi}{\xi} L_i
- y_i \rho g \sin \alpha.
\end{equation}

\noindent
The dynamics of the system is generated by iteratively moving the node with the
minimal value of $T_i$, an infinitesimal distance towards the fluid phase, in
the perpendicular direction to the interface. At each step we need to
accumulate new beads from the initial distribution to the front. This can be
achieved by adding the gathered bead mass which corresponds to the
infinitesimal displacement, to the link cell of the node. If this cell reaches
the threshold value, a new link cell will be assigned, and the rest mass will
be distributed there. This approach will make sure the bead mass field is
conserved.  The chain is interpolated with new nodes as the interface grows,
keeping the resolution of the representation of the interface constant, and the
local quantities, $\kappa_i$ and $L_i$, are recalculated in a neighborhood
along the chain near the moving node.

Note that there is no time in this numerical approach. We can, however,
estimate the time from the volume of the air phase, as we know that the
drainage rate $q$ is constant. This allows us to compare the experimental
results to the numerical simulation during the evolution of the patterns. The
dynamics is deterministic, and the random behavior is a result of perturbed
initial conditions, and imposed quenched fluctuations in the initial mass
field. Note that the random fluctuations in the mass field will induce
fluctuations in $L_i$, as mass is accumulated. These fluctuations scale with
$\sqrt{L_i}$ as $L_i$ correspond to a sum of multiple randomly distributed
masses. This effectively induces fluctuations in $T_i$ evaluated at each node.

We use $\sigma_\xi/\xi=16$~kPa/m, which is an estimate based on comparison
between experimental results and the theoretical expression for finger width
\cite{jon2015numerical}. For the effective surface tension we use
$\gamma=60$~mN/m \cite{knudsen2008pre}. The similarity between the simulated
and experimentally observed patterns (Fig.~\ref{fig:phasediagram}) validates
our theoretical understanding. A noticeable difference between simulations and
experiments is that liquid pathways in the front may break, resulting in
isolated pockets of liquids in the experiments at high $\phi$/low
$\alpha$. These effects are not accounted for in the simulation, as we locally
only track the interface and the grains, not the fluids.

\subsection{Transition of alignment direction}

To understand the transition between horizontally and vertically oriented
finger behavior, we need first to quantify the variations in the yield pressure
threshold [Eq.~(\ref{eq:threshold})]. It is hard to quantify the exact
numerical value of these variations, but it will suffice for our purposes to
determine how the variations scale with $L$. We will assume that these
variations are dominated by the variation of the static friction along the
front [Eq.~(\ref{eq:sigmay})]. If we interpret Eq.~(\ref{eq:sigmay}) to be a
sum of force bearing arc chains of length $\xi$, each of which contributes with
a varying yield stress with a mean value of $\sigma_\xi$, then the total
variation will scale with the number of these chains. As the number of chains
scales with the size of the front, we have that Var$(\sigma_Y) \propto L$, and
that the standard deviation is proportional to $\sqrt L$. Note that in the
numerical simulations the value of $\sigma_\xi$ is kept fixed (it is not a
random variable), but we introduce another physical source of fluctuation
leading to the same scaling. We impose the fluctuations in the initial bead
density field, which induces fluctuations in the front thickness that also
scale with $\sqrt L$ for a fixed displacement of the interface.

We can compare these variations to the hydrostatic pressure difference over a
horizontally oriented finger. The finger width is $2\Lambda$, and the
corresponding hydrostatic difference is $2\Lambda g \rho \sin \alpha$. The
ratio between the standard deviation of the yield stress, and the hydrostatic
difference of a horizontally oriented finger is therefore,

\begin{equation}
  \eta \propto \frac{\sqrt{L}}{\Lambda \sin \alpha}.
\label{eq:eta0}
\end{equation}

\noindent
This ratio indicates the behavior of the alignment. When the contribution of
stress fluctuations is comparable to the stabilizing pressure ($\eta\simeq1$) a
finger can get ahead of its neighbors and grow sideways, orthogonal to the
direction of gravity. For $\eta<1$, the fluctuations fail to disrupt the
side-by-side finger growth. For $\eta>1$, the fluctuations dominate over the
stabilizing effect, and the alignment is lost.  We can only estimate $\eta$ up
to a multiplicative constant, as the numerical value of the stress variations
of $\sigma_Y$ is hard to identify. This will, however, suffice for identifying the
contour lines in the $(\alpha,\phi)$ plane, which have similar alignment properties. To
identify these contour lines, we first need to express $\Lambda$ and $L$ in
terms of $\phi$.

Let $A$ and $C$ be respectively the area and the circumference of the air
phase, as seen from above, and let $h$ be the cell gap. The pattern is
dominated by finger structures, such that $A = C \Lambda$. We assume that $L$
is approximately constant along the interface, such that $CL$ is the total area
of the front. Mass conservation gives that $h(CL+A)\phi = hCL$, which under the
substitution $\Lambda = A/C$, implies that

\begin{equation}
  L=\Lambda
 \,\frac{\phi}{1-\phi}.
\label{eq:L}
\end{equation}

\noindent
A more detailed derivation, which differentiates between the front thickness at
the sides and the tip of the fingers, yields correction terms to this
expression (see Ref.~\cite{jon2015numerical}).

The work of a typical displacement, $\delta w$, has two contributions when we
set $\alpha=0$ for simplicity. First, the stretching of the interface
contributes with $\gamma h \,\delta C$, where $\delta C = \delta A / \Lambda$,
which follows from the assumption of constant $\Lambda$. Second, the work done
against the granular stresses, $\sigma$, in the front, is
$h s \, \delta x \,\sigma$, where $s$ is the typical width of a moving segment
and $\delta x$ is the distance the interface advances such that
$s \delta x = \delta A$.  We can approximate $\sigma$, by the yield stress,
$\sigma_Y$ [Eq.~(\ref{eq:sigmay})].  Putting the terms together, and dividing
by the displacement duration, gives the work rate,

\begin{equation}
  \frac{\delta w}{\delta t} = \left(\frac{\gamma }{\Lambda} +  L \frac{\sigma_\xi}{\xi}\right) h \frac{\delta A}{\delta t},
  \label{eq:dissipation}
\end{equation}

\noindent
where $h \delta A/\delta t$ equals the constant compression rate, when averaged
over many stick-slip events. Substituting Eq.~(\ref{eq:L}) and
minimizing Eq.~(\ref{eq:dissipation}) with respect
to $\Lambda$ gives

\begin{equation}
0 = \frac{\mathrm{d}}{\mathrm{d} \Lambda}\frac{\delta w}{\delta t} 
\, \Rightarrow \, 0 =\frac{\mathrm{d}}{\mathrm{d}\Lambda} \left(\frac{\gamma }{\Lambda} +  \Lambda \frac{\phi}{1-\phi} \frac{\sigma_\xi}{\xi}\right),
\end{equation}

\noindent
which corresponds to the assumption that the pattern evolves in a
way that minimizes the work. This implies 

\begin{equation}
   \Lambda \propto \sqrt{\frac{1-\phi}{\phi}}.
   \label{eq:Lambda}
\end{equation}

\noindent
We can now use Eqs.~(\ref{eq:L}) and (\ref{eq:Lambda}) to rewrite
Eq.~(\ref{eq:eta0}) as a function of $\phi$ and $\alpha$,

\begin{equation}
  \eta \propto\frac{1}{\sin\alpha}\left(\frac{\phi}{1-\phi}\right)^{3/4}.
  \label{eq:eta}
\end{equation}

\noindent
Indeed, contours of constant $\eta$ correspond to equal qualitative alignment behavior,
as shown in Fig.~\ref{fig:phasediagram}.

\section{Application: Flow in dykes}

\begin{figure}
  \includegraphics[width=8.6cm]{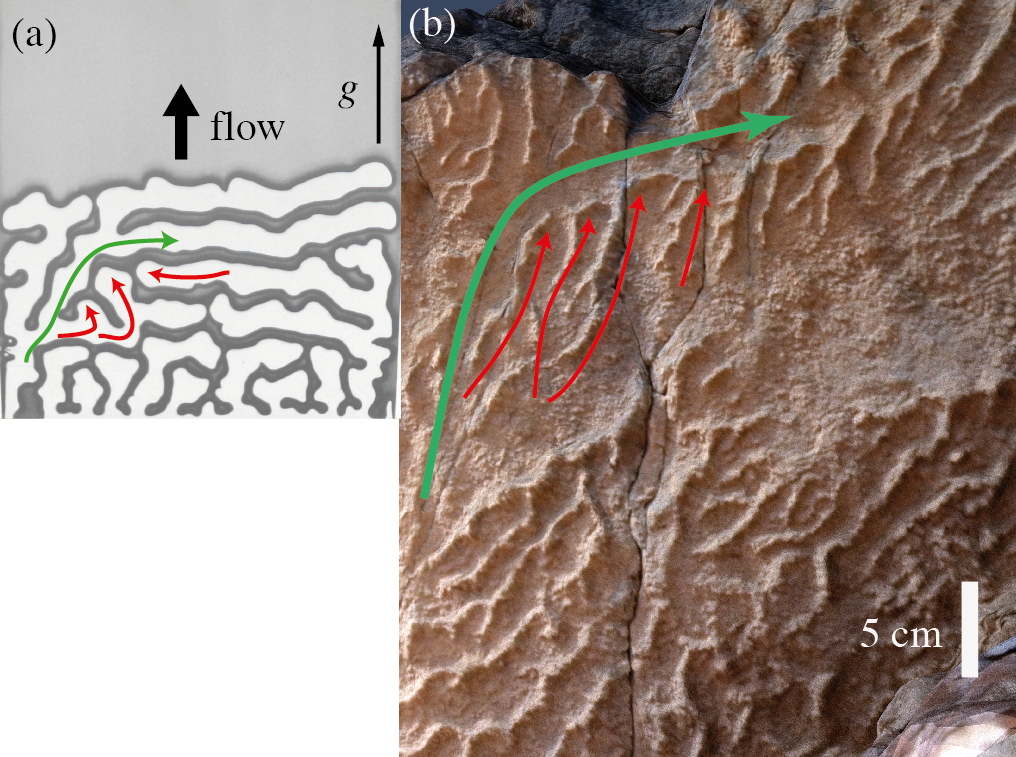}
  \caption{(Color online) Feature comparison 1 between the experimental
    observations at $\phi=0.4$, $\alpha=4^\circ$ (a) remaining structures on
    dyke walls found in the Inmar formation (b). Fingers (red arrows) are being
    intercepted by a finger (green arrows) which grows perpendicular to the
    average flow direction. The gravitational pull is indicated by $g$. The
    scale bar in (b) applies to both experiment and dyke figure.}
\label{fig:feature1}
\end{figure}

\begin{figure}
  \includegraphics[width=8.6cm]{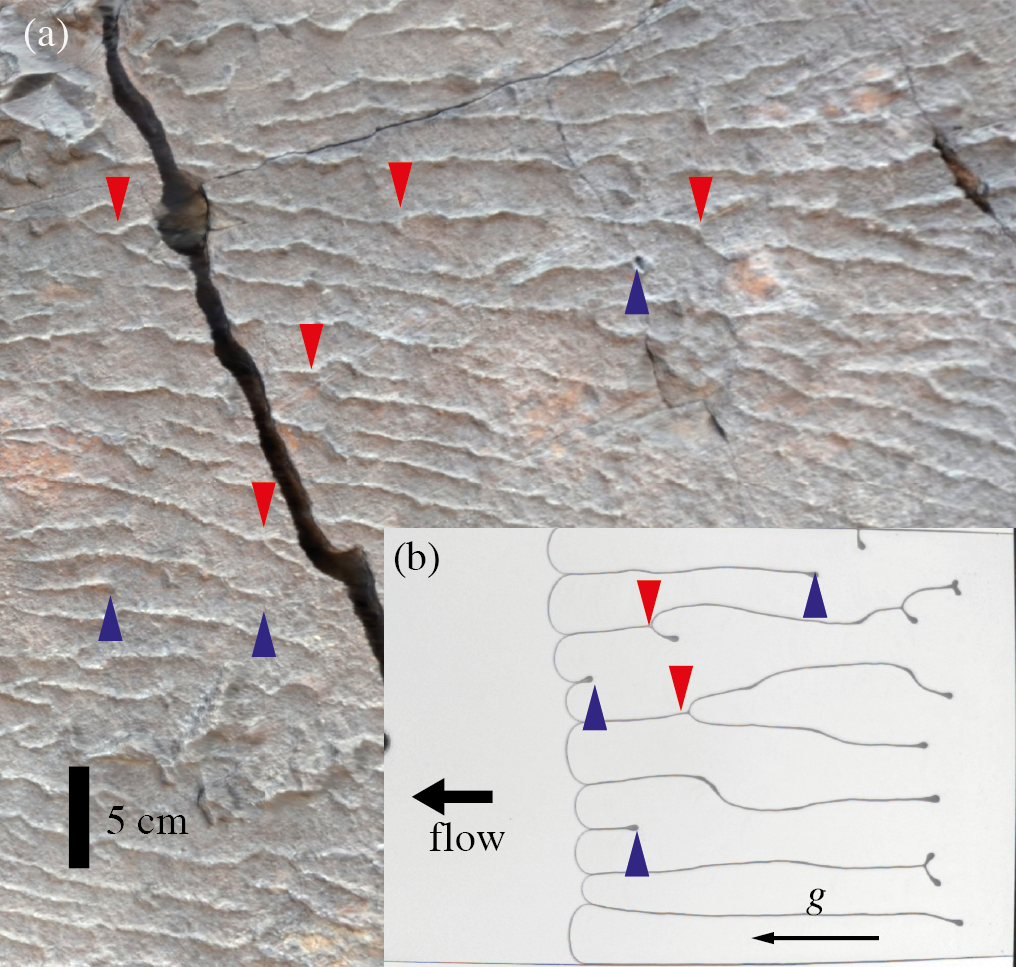}
  \caption{(Color online) Feature comparison 2 between the experimental
    observations at $\phi=0.025$, $\alpha=4^\circ$ (b) and the remaining
    structures on dyke walls found in the Inmar formation (a). Aligned finger
    structures with tip-splitting and termination, respectively marked by blue
    and red triangles. The gravitational pull is indicated by $g$. The scale
    bar in (a) applies to both experiment and dyke figure.}
\label{fig:feature2}
\end{figure}

We now turn to the relevance of this system to magmatic flow during dyke
formations. A (magmatic) \emph{dyke} is an approximately two-dimensional sheet-like body of magma, which has penetrated a pre-existing body of rock in a
direction which is perpendicular to the bedding planes (i.e.~the planes of the
sediments).  A striking example of a dyke formation is found in the Inmar
formation in the desert in southern Israel.  There, the igneous rock (i.e. the
solidified magma) of the dyke has eroded away, and the erosion resistant dyke
walls, made by quartzitic sandstone, are exposed. These walls display a rich
network of finger structures~\cite{baer1987, baer1991}. The fingers are
identified as grooves in the sandstone; outward bulging ridges separate the
fingers from its neighbors. A finger is approximately 1-10~cm wide and
10-100~cm long, and the wall shows intermittent patches of finger alignment
[Figs.~\ref{fig:feature1} (b) and~\ref{fig:feature2} (a)]. The walls are
separated $\simeq1$~m apart, but mirror images of the structures remain on both
walls, which suggests that the structures were made during the initial stages
of the dyke formation.

The ridges contain a closely packed concentration of quartz grains
(100-500~$\upmu$m diameter) cemented by iron oxides and kaolinite, in contrast
to the relatively low concentration of quartz grains (similar in composition) in the rock near the
grooves~(see Fig.~12 in Ref.~\cite{baer1991}).  The finger structures indicate
that the sandstone was fluidized during the formation while these quartz grains were preserved as solid, and that the grains were
accumulated onto stagnant regions adjacent to the interface of the invading
magma, which initially filled the grooves.

The finger formation in these dykes has previously been explained as viscous
fingers due to the potential viscous contrast between an invading dyke-related
fluid and the fluidized host rock~\cite{baer1991}. Viscous fingers in porous
media are, however, known to display fractal invasion patterns with no
intrinsic length scale \cite{maaloy1985prl,lenormand1985,chen1985}, whereas the
fingers on the dyke walls display a characteristic width. The similarity of
these dyke wall fingers to the aligned finger structures observed in our
experimental setup, suggests that the fingers are generated by inter granular
friction between the quartz grains and accumulation of these grains onto
stagnant fronts.

The relevance of our system to the structure in the Inmar formation is further
substantiated by the similarity in the features of the resulting pattern. In
particular we observe similar tip-splitting and termination properties
[Fig.~\ref{fig:feature2}], and interception of fingers by a finger which grows
perpendicular to the average flow direction [Fig.~\ref{fig:feature1}].

The fingers direction in the Inmar formation varies locally between vertically
upwards and downwards. Steps in the dyke structures indicate that the intrusion
was following a propagating crack~\cite{baer1991}. The consistent direction of the crack direction, and the vertical orientation of the dykes, indicate that no large geological deformation took place after the formation. The direction of the gravitational effect on the fingers, depends on the density contrast of the invading fluid to the fluidized host rock, which is unknown. Variations in the crack
opening will, however, induce a stabilizing potential in the capillary pressure as the
out-of-plane component of the magma interface curvature increases towards the
crack tip. A combination of hydrostatic and capillary pressure variations, is
therefore likely to act as the stabilizing potential in the dyke finger
formation. Variations in the crack spacing also explain the local variations in
the finger directions, and the presence of features [Figs.~\ref{fig:feature1}
(b) and ~\ref{fig:feature2} (a)], from different parts of the phase diagram
[Fig.~\ref{fig:phasediagram} (c)].

\section{Conclusion}

We have described a new type of pattern forming flow, where
grains are accumulated by a moving interface, which, when subject to a
stabilizing potential, forms aligned finger structures. We identify the finger
width by a work minimization principle, and can estimate the alignment
direction by the competition between frictional force fluctuations and the
hydrostatic pressure.  The dynamics is quasi-static; it depends on granular
friction rather than viscosity. The  patterning process seems to be independent
of whether the invading fluid is a gas or a liquid, as long as the phases are
immiscible. We can reproduce the finger behavior numerically by accounting for
the hydrostatic pressure, grain accumulation, solid friction and capillary
forces. As our model only contains geologically ubiquitous mechanisms, it may
be relevant for a number of biphasic flow phenomena confined to planar
fractures, in particular multiphase flow during dyke formation that leave
imprints of the finger formation as solidified granular residue on the dyke
walls.

\begin{acknowledgments}
  We thank the late Henning Knudsen, who made important contributions to the
  understanding of frictional fingers. We thank Gidon Baer, Einat Aharonov,
  Olivier Galland and Benjy Marks for discussions. J.A.E.~acknowledges support
  from the Research Council of Norway through the NFR Project No.~200041/S60,
  the Campus France Eiffel Grant and Unistra. B.S.~acknowledges support from
  the EPSRC Grant EP/L013177/1 and S{\^e}r Cymru National Research Network in Advanced Materials Grant no. NRN141. R.T.,~K.J. and E.G.F.~acknowledges support from
  EU's FP7 grant no.~316889-ITN FlowTrans. R.T.~also acknowledges additional
  support from UiO, Unistra and the INSU risk program.
\end{acknowledgments}

\bibliography{refs}
\end{document}